# Spectroscopy of flux-driven Kerr parametric oscillators by reflection coefficient measurement

Aiko Yamaguchi[1,2*], Shumpei Masuda[2,3], Yuichiro Matsuzaki[4], Tomohiro Yamaji[1,2],

Tetsuro Satoh[1,2], Ayuka Morioka[1,2], Yohei Kawakami[1,2], Yuichi Igarashi[1,2],

Masayuki Shirane[1,2], and Tsuyoshi Yamamoto[1,2]

[1]Secure System Platform Research Laboratories, NEC Corporation,

Kawasaki, Kanagawa 211-0011, Japan

[2]NEC-AIST Quantum Technology Cooperative Research Laboratory,

National Institute of Advanced Industrial Science and Technology (AIST),

Tsukuba, Ibaraki 305-8568, Japan

[3]Global Research and Development Center for Business by

Quantum-AI Technology (G-QuAT), AIST,

National Institute of Advanced Industrial Science and Technology (AIST),

1-1-1, Umezono, Tsukuba, Ibaraki 305-8568, Japan

[4]Department of Electrical, Electronic, and Communication Engineering,

Faculty of Science and Engineering, Chuo University,

1-13-27 Kasuga, Bunkyo-ku, Tokyo 112-8551, Japan.

We report the spectroscopic characterization of a Kerr parametric oscillator (KPO) based on the measurement of its reflection coefficient under a two-photon drive induced by flux modulation. The measured reflection spectra show good agreement with numerical simulations in terms of their dependence on the two-photon drive amplitude. The spectra can be interpreted as changes in system's eigenenergies, transition matrix elements, and the population of the eigenstates, although the linewidth of the resonance structure is not fully explained. We also show that the drive-amplitude dependence of the spectra can be explained analytically by using the concepts of Rabi splitting and the Stark shift. By comparing the experimentally obtained spectra with theory, we show that the two-photon drive amplitude at the device can be precisely determined, which is important for the application of KPOs in quantum information processing.

I. Introduction

Dressed states formed in a driven quantum system are typically characterized by spectroscopic measurements. For example, if a two-level system is driven on resonance, the Mollow triplet [1] is observed due to the Rabi splitting of the two energy levels hybridized with the photon number states. If, on the other hand, the drive is far detuned, the shift of the energy level due to the Autler-Townes effect, namely, the ac Stark effect [2] is observed.

The experiments originally done in atomic system have not only been reproduced, but also extended to an unexplored parameter regime using superconducting artificial atoms, referred to as a circuit quantum electrodynamics (c-QED) system [3,4]. For example, the ac Stark shift in a c-QED system was first reported in Ref. [5], and extended to the strong dispersive regime, where it is possible to determine the population of each photon number state in the resonator coupled to the qubit [6].

Compared to the experiments using one-photon drive, spectroscopy of a two-photon driven system has been relatively unexplored in c-QED systems. The two-photon drive plays an essential role in a device called a Kerr parametric oscillator (KPO) [7,8], which has been attracting attention for applications in quantum information processing, such as quantum annealing [9,10] and gate-model quantum computing [11–13]. KPOs are parametric oscillators, in which the Kerr nonlinearity is larger than the photon loss rate, particularly in the single-photon Kerr regime [14]. The KPOs can be implemented using Josephson parametric oscillators [15–19] or charge-driven transmons with a superconducting nonlinear asymmetric inductive element (SNAIL) [20–22]. In both cases, an external field nearly twice the resonance frequency of the KPOs is used for parametric pumping and works as the two-photon drive.

Spectroscopic measurement for studying the energy-level structure up to the tenth excited states of the KPO of charge-driven transmons has been reported by Frattini et al. [21]. By measuring the excitation energy from the ground states as a function of the two-photon drive amplitude, they revealed the crossover from a nondegenerate spectrum to a pairwise kissing spectrum formed by two excited states in the double-well meta potential of the KPO, which leads to a staircase pattern in the coherent state lifetime. This type of spectroscopic measurement for revealing the energy spectra of KPOs is important not only for understanding its fundamental physics, but also for its application to quantum information processing. For example, in Ref. [23], a gate operation of the cat qubit using excited energy levels was proposed. Moreover, a comparison of the spectroscopic measurement with the theory generally gives the calibration of the drive amplitude at the quantum chip [5,24,25]. This is particularly important in KPOs for precisely determining the two-photon drive amplitude to create an optimal schedule to generate Schrödinger's cat state in a short time [13,26,27]

Some of the authors of the present study theoretically proposed an alternative method of spectroscopy for revealing the energy level structure of a KPO under two-photon drive [28]. The method is based on the measurement of reflection coefficient of the KPO and is simpler than that used in previous studies [21,22] because it is only performed by continuous wave measurement without using microwave pulses. In the present paper, we experimentally demonstrate the proposed method by performing reflectometry measurements of the KPO under the two-photon drive induced by the flux modulation. We compare the results with numerical simulations on the basis of the theory to interpret the experimental data. We also show that the experimental data can be used to determine the calibration of the two-photon drive amplitude, which is important for the application of KPOs in quantum information processing.

## II. Device

The KPO used in this experiment is a lumped-element-type device. An optical microscope image and an equivalent circuit of the KPO are shown in Figs. 1(a) and 1(b), respectively. The left end face of the KPO is connected to a signal line via a coupling capacitor $C_{\text{in}}$, from which a probe signal is applied. The bottom end face is connected to a coupler for interaction with other KPOs, which are not used in this experiment. The top end face is connected to the ground plane via two Josephson junctions (JJs) and a SQUID in series, the latter of which is inductively coupled to the pump line. The JJ is placed in series with the SQUID to reduce the magnitude of the Kerr nonlinearity $\chi$ of the KPO [29]. The SQUID contains asymmetric JJs, whose critical currents are designed to be 375 nA and 625 nA, and those of the series junctions are designed to both be 1000 nA. The actual critical current of each JJ was verified by measuring the room-temperature resistance of a test structure. The KPO is placed in a dilution refrigerator and cooled to below 10 mK.

We measure the reflection coefficient of the KPO without the two-photon drive to extract the parameters of the KPO. The reflection coefficient $\Gamma$ is expressed by the following equation,

$$\Gamma = 1 - \frac{\kappa_e}{i(\omega_{\text{ref}} - \omega_r) + \frac{\kappa_e + \kappa_i}{2}}, \qquad (1)$$

where $\omega_r$ is the resonance frequency of the KPO, $\omega_{\text{ref}}$ is the probe frequency, and $\kappa_e$ ($\kappa_i$) is an external (internal) loss rate. Figure 1(c) shows the flux-bias dependence of the resonance frequency. The resonance frequency of the KPO changes periodically with the flux bias induced by the DC current applied to the pump line. In the following measurement, we fix the DC flux bias to $0.385\Phi_0$, where $\Phi_0$ is the magnetic flux quantum. At this flux bias, the resonance frequency $\omega_r/2\pi = 8.9653$ GHz, the external loss rate $\kappa_e/2\pi = 0.27$ MHz and the internal loss rate $\kappa_i/2\pi = 0.45$ MHz are estimated by fitting the measured $\Gamma$ to Eq. (1). Note that the internal loss rate obtained from this reflection measurement may include the contribution from pure dephasing; the actual internal loss rate $\kappa_i^*$ is related to $\kappa_i$ as $\kappa_i^* = \kappa_i - 2\gamma$, where $\gamma$ is the pure dephasing rate of the KPO.

### III. Measurement of reflection coefficient under two-photon drive

Next, we measure $\Gamma$ under the two-photon drive using a vector network analyzer (VNA). The two-photon drive generated by an additional microwave source at room temperature is applied to the pump line on the chip [Fig. 1(b)]. The frequency of the two-photon drive $\omega_p$ is approximately twice the resonance frequency of the KPO, and the detuning is defined as the difference between the resonance frequency and half of the two-photon-drive frequency, that is, $\Delta = \omega_r - \omega_p/2$.

The Hamiltonian of the KPO in a frame rotating at $\omega_p/2$ is expressed as [16]

$$\frac{\mathcal{H}}{\hbar} = \Delta a^\dagger a - \frac{\chi}{2} a^\dagger a^\dagger a a + \beta(a^\dagger a^\dagger + aa), \qquad (2)$$

where $a$ is the photon annihilation operator, and $\beta$ is the amplitude of the two-photon drive. The Kerr nonlinearity $\chi/2\pi = 17.0$ MHz is determined by measuring the probe-power dependence of $\Gamma$ [15]. The Fock state $|n\rangle$, where $n$ is the photon number, is the eigenstate with eigenenergy of $n\Delta - n(n-1)\chi/2$ when the two-photon drive is absent. To examine situations with varying degeneracy of eigenstates, we use three different $\Delta$'s ; $\Delta = 2\pi \times 8.20$ MHz $(\sim \chi/2)$, $2\pi \times 0.05$ MHz $(< \kappa_{total})$, and $2\pi \times (-8.10)$ MHz $(\sim -\chi/2)$, where $\kappa_{total}(= \kappa_e + \kappa_i)$ is the total photon loss rate. If $\Delta = \chi/2$ ($\Delta = 0$), the Fock states $|0\rangle$ and $|2\rangle$ ($|1\rangle$) are degenerate at $\beta = 0$. In the case of $\Delta = -\chi/2$, there are no degenerate states at $\beta = 0$.

Figure 2 shows the amplitude of $\Gamma$ as a function of $\omega_{ref}$ and the two-photon drive power at the generator output $\bar{p}_{RT}$ for different detunings [(a) $\Delta/2\pi = +8.20$ MHz, (b) $+0.05$ MHz, (c) $-8.10$ MHz (c)]. Here, $\bar{p}_{RT}$ was changed for each scan of $\omega_{ref}$ using the VNA. In all cases, we observed single dips in $|\Gamma|$ corresponding to the transition from the state $|0\rangle$ to the state $|1\rangle$ for sufficiently small $\bar{p}_{RT}$. As $\bar{p}_{RT}$ is increased, the spectra exhibits other transitions, and the patterns and the number of transitions [6, 2, and 3 for (a), (b), and (c), respectively] vary with the detuning. In addition, amplification peaks ($|\Gamma| > 1$) are observed in Figs. 2(a) and 2(c), while they are absent in Fig. 2(b). For $\Delta/2\pi = +8.20$ MHz, the transition observed at $(\omega_{\text{ref}} - \omega_{\text{p}}/2)/2\pi = +8.20$ MHz is an absorption dip at small $\bar{p}_{RT}$ and splits into two amplification peaks as $\bar{p}_{RT}$ is increased. In the region of $\bar{p}_{RT} > -30$ dBm, three of the six transitions shift toward higher frequency, and the others toward lower frequency as $\bar{p}_{RT}$ is increased. For $\Delta/2\pi = +0.05$ MHz, two transitions are observed, and the frequency of the transition observed at $(\omega_{\text{ref}} - \omega_{\text{p}}/2)/2\pi = 0.05$ MHz is constant and that of the other transition becomes lower in frequency as $\bar{p}_{RT}$ is increased. For $\Delta/2\pi = -8.10$ MHz, one of the three transitions is observed as an amplification peak, and the others are absorption dips. One of the absorption dips shifts toward higher frequency, and the other toward lower frequency as $\bar{p}_{RT}$ is increased.

## IV. Numerical simulations

We numerically simulate the reflection coefficient for comparison with the experiments. The reflection coefficients are calculated following Ref. [28] as

$$\Gamma = 1 + \sum_{\widetilde{m},\widetilde{n}} \xi_{\widetilde{m}\widetilde{n}} \qquad (3)$$

with

$$\xi_{\widetilde{m}\widetilde{n}} = \frac{\kappa_{\text{e}}|X_{\widetilde{m}\widetilde{n}}|^2(\rho_{\widetilde{m}\widetilde{m}} - \rho_{\widetilde{n}\widetilde{n}})}{i(\omega_{\text{ref}} - \omega_{\widetilde{m}\widetilde{n}}) + \frac{\kappa_{\text{e}} + \kappa_{\text{i}}}{2}(Y_{\widetilde{m}\widetilde{m}} + Y_{\widetilde{n}\widetilde{n}})}. \qquad (4)$$

Here, $\xi_{\widetilde{m}\widetilde{n}}$ is the contribution of the transition from the state $|\widetilde{n}\rangle$ to the state $|\widetilde{m}\rangle$ in the reflection coefficient. The state $|\widetilde{n}\rangle$ is the energy eigenstate of the KPO, and the state is labeled by the photon number of the state when the two-photon drive is absent. Here, $\omega_{\widetilde{m}\widetilde{n}}$ is the transition frequency from the state $|\widetilde{n}\rangle$ to the state $|\widetilde{m}\rangle$ calculated by diagonalizing the Hamiltonian Eq. (2). $\rho_{\widetilde{n}\widetilde{n}}$ is the $\widetilde{n}$-th diagonal element $\langle\widetilde{n}|\rho|\widetilde{n}\rangle$ of the steady-state density matrix obtained by solving a GKSL master equation with a single-photon loss, $X_{\widetilde{m}\widetilde{n}} =$ $\langle \widetilde{m}|a^\dagger|\widetilde{n}\rangle$ is the matrix element of the one-photon transition, and $Y_{\widetilde{m}\widetilde{m}} = \langle \widetilde{m}|a^\dagger a|\widetilde{m}\rangle$ is the expectation value of the photon number for the state $|\widetilde{m}\rangle$.

Figures 3(a)-3(c) show the result of the numerical simulations, which are qualitatively in agreement with the experimental results shown in Figs. 2(a)-2(c). For quantitative comparison in the following sections, here we convert the horizontal axes in Fig. 2, the power of the two-photon drive $\bar{p}_{RT}$, into the amplitude of the two-photon drive $\beta$, which can be calculated by the first-order derivative of resonance frequency with respect to the current applied to the pump line $d\omega_r/di$ (Appendix A):

$$\frac{\beta}{2\pi} = \frac{1}{4}\left|\frac{d\omega}{di}\right|\sqrt{\frac{2}{Z_0 \times 1000}} 10^{\frac{\bar{p}_{KPO}}{20}}. \tag{5}$$

From the measurement results shown in Fig. 1. (c), we obtained $d\omega/di = 2\pi \times (-8.27)$ MHz/$\mu$A at this bias point. Another parameter, the two-photon drive power at the chip $\bar{p}_{KPO}$, is $\bar{p}_{KPO} = \bar{p}_{RT} + R_{RT-KPO}$, where $R_{RT-KPO}$ is the attenuation of the two-photon drive inside and outside the refrigerator. The attenuation $R_{RT-KPO}$ is estimated by comparing the measured transition frequencies with the theoretical ones, which were calculated using the Hamiltonian in the absence of a coherent drive under the rotating wave approximation. If a transition frequency is separated from others by a frequency larger than the line width, the transition frequency is extracted from the measured spectrum [Fig. 2(a)-(c)] by fitting it to the following equation [13]

$$\Gamma_{\widetilde{m}\widetilde{n}} = 1 - \frac{\kappa_e^{(\widetilde{m}\widetilde{n})}}{i(\omega_{\text{ref}} - \omega_{\widetilde{m}\widetilde{n}}) + \frac{\kappa_e^{(\widetilde{m}\widetilde{n})} + \kappa_i^{(\widetilde{m}\widetilde{n})}}{2}}, \tag{6}$$

where $\kappa_e^{(\widetilde{m}\widetilde{n})}$ and $\kappa_i^{(\widetilde{m}\widetilde{n})}$ correspond to nominal external and internal loss rates, respectively. To extract $R_{RT-KPO}$, we used $R_{RT-KPO}$ as a fitting parameter minimizing the square of the difference between the measured and calculated transition frequencies. The estimated values of $R_{RT-KPO}$ are -57.0 dB, -57.6 dB, and -57.6 dB for $\Delta/2\pi = +8.20$ MHz, $+0.05$ MHz, and $-8.10$ MHz, respectively. From an independent measurement using a chip with a through transmission line, we estimate that the two-photon drive is attenuated by 58 dB from the microwave source to the KPO with an accuracy of $\pm 2$ dB, which is consistent with the above result.

Figures 3(d)-3(f) show the same experimental data as Fig. 2 plotted as a function of $\beta$ in the horizontal axis. The theoretical calculation of the transition frequencies $\omega_{\widetilde{m}\widetilde{n}}$'s is also plotted by gray dashed lines, which very well reproduces the observed transitions.

## V. Interpretation of transition spectra

The experimental results [Figs. 3(d)-3(f)] are highly consistent with the numerical calculations [Figs. 3(a)-3(c)]. The agreement indicates that the theoretically predicted energy levels and the population distribution of the stationary state are reproduced in the experiment. From here, we interpret each transition spectrum shown in Figs. 3(a)-3(c). Note that the two-photon drive does not change the photon number parity, whereas the one-photon drive used as the probe signal does. Therefore, transitions can occur between the states $|\widetilde{2k}\rangle$ and $|\widetilde{(2k'+1)}\rangle$, where both $k$ and $k'$ are non-negative integers. For the transitions $|\widetilde{n}\rangle \to |\widetilde{m}\rangle$ at $\omega_{\widetilde{m}\widetilde{n}}$ to be visible, both $|X_{\widetilde{m}\widetilde{n}}|$ and $(\rho_{\widetilde{m}\widetilde{m}} - \rho_{\widetilde{n}\widetilde{n}})$ must be non-zero [Eq. (4)]. Whether the transition at $\omega_{\widetilde{m}\widetilde{n}}$ appears as a peak or dip in the spectrum is determined by the sign of the population difference $(\rho_{\widetilde{m}\widetilde{m}} - \rho_{\widetilde{n}\widetilde{n}})$. Thus, the reflection measurement provides information about the eigenenergy, population distribution, and transition matrix element.

First, we consider $\Delta/2\pi = 8.20$ MHz. Figures 3(g) and 3(j) show the numerically calculated energy levels and their populations, respectively, for the four highest energy eigenstates of the KPO in the rotating frame as a function of $\beta$. The energy level and the state population are schematically drawn in Figs. 4(a) and 4(c). In the case of $\beta/2\pi = 0.21$ MHz $\ll \chi/2\pi$ [Fig.4(a)], the states $|\widetilde{0}\rangle$ and $|\widetilde{2}\rangle$ are almost degenerate. Because $\beta$ is sufficiently smaller than $\kappa_{\text{total}}$, the population is concentrated in the state $|\widetilde{0}\rangle$. Therefore, the transition $|\widetilde{0}\rangle \to |\widetilde{1}\rangle$ ($|\widetilde{1}\rangle \to |\widetilde{0}\rangle$) should appear as an absorption dip (amplification peak). Figure 4(b)

plots both experimentally obtained and numerically calculated $\Gamma$'s as a function of $\omega_{\text{ref}}$. As expected, we observe an absorption dip at $\omega_{\tilde{1}\tilde{0}}$ and a small amplification peak at $\omega_{\tilde{0}\tilde{1}}$ and their difference in magnitude comes from the difference in their transition matrix elements of the one-photon drive. Similarly, the transitions at $\omega_{\tilde{2}\tilde{1}}$ and $\omega_{\tilde{3}\tilde{0}}$ appear as absorption dips. The fact that the transition between the states $|\tilde{0}\rangle$ and $|\tilde{3}\rangle$ exhibit weak probe power indicates that at least one of them deviates from the Fock states.

Figures 4(c) and 4(d) show the schematic energy diagram and the reflection coefficient at $\beta/2\pi = 7.40$ MHz, respectively. Because the population inversion ($\rho_{\tilde{0}\tilde{0}} < \rho_{\tilde{1}\tilde{1}}$) is created, the transition $|\tilde{0}\rangle \to |\tilde{1}\rangle$ at $\omega_{\tilde{1}\tilde{0}}$ turns out to be an amplification peak, which is referred to as dressed-state amplification [30]. Because the degeneracy of the states $|\tilde{0}\rangle$ and $|\tilde{2}\rangle$ is lifted at $\beta/2\pi = 7.40$ MHz [Fig. 3(g)], the transition $|\tilde{2}\rangle \to |\tilde{1}\rangle$ is separated from the transition $|\tilde{0}\rangle \to |\tilde{1}\rangle$ and becomes visible as another amplification peak at $\omega_{\tilde{1}\tilde{2}}$. The transition $|\tilde{2}\rangle \to |\tilde{3}\rangle$ is separated from the transition $|\tilde{0}\rangle \to |\tilde{3}\rangle$ and also becomes visible as another dip at $\omega_{\tilde{3}\tilde{2}}$. Because four states $|\tilde{0}\rangle$, $|\tilde{1}\rangle$, $|\tilde{2}\rangle$, and $|\tilde{3}\rangle$ are involved, there are eight possible transitions $|\tilde{0}\rangle \leftrightarrow |\tilde{1}\rangle$, $|\tilde{0}\rangle \leftrightarrow |\tilde{3}\rangle$, $|\tilde{2}\rangle \leftrightarrow |\tilde{1}\rangle$, and $|\tilde{2}\rangle \leftrightarrow |\tilde{3}\rangle$, which appear as four absorption dips and four amplification peaks. However, the transitions $|\tilde{3}\rangle \to |\tilde{0}\rangle$ and $|\tilde{3}\rangle \to |\tilde{2}\rangle$ cannot be observed because the matrix element of the one-photon transition $|X_{\tilde{0}\tilde{3}}|$ and $|X_{\tilde{2}\tilde{3}}|$ are too small as shown in Fig. 3(m). This makes the number of transitions six.

For $\beta/2\pi > 10$ MHz, some of the transitions disappear as shown in Fig. 3(a). The peak and dip in the transitions $|\tilde{0}\rangle \to |\tilde{1}\rangle$ at $\omega_{\tilde{1}\tilde{0}}$ and $|\tilde{1}\rangle \to |\tilde{0}\rangle$ at $\omega_{\tilde{0}\tilde{1}}$ disappear because the population difference between $\rho_{\tilde{0}\tilde{0}}$ and $\rho_{\tilde{1}\tilde{1}}$ is small [Fig. 3(j)]. Similarly, the dip in the transition $|\tilde{2}\rangle \to |\tilde{3}\rangle$ at $\omega_{\tilde{3}\tilde{2}}$ disappears because the difference between $\rho_{\tilde{2}\tilde{2}}$ and $\rho_{\tilde{3}\tilde{3}}$ is small. The peak height of the transition $|\tilde{2}\rangle \to |\tilde{1}\rangle$ at $\omega_{\tilde{1}\tilde{2}}$ becomes small due to decreasing $|X_{12}|$ [Fig. 3(m)].

In the case of $\Delta/2\pi = +0.05$ MHz, the states $|\tilde{0}\rangle$ and $|\tilde{1}\rangle$ are almost degenerate regardless of the value of $\beta$ as shown in Fig. 3(h). Within the probe frequency range in this measurement [Fig. 3(e)], the transitions $|\tilde{0}\rangle \leftrightarrow |\tilde{1}\rangle$ and $|\tilde{1}\rangle \leftrightarrow |\tilde{2}\rangle$ can contribute to the spectrum. However, the transition $|\tilde{2}\rangle \rightarrow |\tilde{1}\rangle$ at $\omega_{\tilde{1}\tilde{2}}$ is not visible because the matrix element of the one-photon transition $|X_{\tilde{1}\tilde{2}}|$ is small [Fig. 3(n)]. For the transitions $|\tilde{0}\rangle \leftrightarrow |\tilde{1}\rangle$, the frequencies $\omega_{\tilde{1}\tilde{0}}$ and $\omega_{\tilde{0}\tilde{1}}$ are the same, so the dip at $\omega_{\tilde{1}\tilde{0}}$ overlaps with the peak at $\omega_{\tilde{0}\tilde{1}}$, resulting in two distinguishable spectra. As the populations of $\rho_{\tilde{0}\tilde{0}}$ and $\rho_{\tilde{1}\tilde{1}}$ approach 0.5 [Fig. 3(k)], the dip depth at $\omega_{\tilde{1}\tilde{0}}$ appears to decrease [Figs. 3(b) and 3(e)]. In contrast, although the matrix element of the one-photon transition $|X_{\tilde{2}\tilde{1}}|$ slightly decreases [Fig. 3(n)], the peak of $|\tilde{1}\rangle \rightarrow |\tilde{2}\rangle$ at $\omega_{\tilde{2}\tilde{1}}$ becomes more visible, because the population difference between $\rho_{\tilde{1}\tilde{1}}$ and $\rho_{\tilde{2}\tilde{2}}$ increases [Fig. 3(k)]. For $\beta \gg \chi$, the steady state of the states $|\tilde{0}\rangle$ and $|\tilde{1}\rangle$ are the same mixed state [10,11].

For $\Delta/2\pi = -8.10$ MHz, there are no degenerate states at $\beta = 0$ [Fig. 3(i)]. We observe one amplification peak at $\omega_{\tilde{0}\tilde{1}}$ and two absorption dips at $\omega_{\tilde{1}\tilde{0}}$ and $\omega_{\tilde{2}\tilde{0}}$, which can be attributed to the relationship between the populations $\rho_{\tilde{0}\tilde{0}} > \rho_{\tilde{1}\tilde{1}} > \rho_{\tilde{2}\tilde{2}}$ [Fig. 3(l)]. As the population $\rho_{\tilde{1}\tilde{1}}$ increases, the transition $|\tilde{1}\rangle \rightarrow |\tilde{2}\rangle$ become more visible as the dip at $\omega_{\tilde{2}\tilde{1}}$ in $|\Gamma|$ [Figs. 3(c) and 3(f)]. In the same regime, although the population difference between $\rho_{\tilde{0}\tilde{0}}$ and $\rho_{\tilde{1}\tilde{1}}$ decreases, the peak at $\omega_{\tilde{0}\tilde{1}}$ become visible because $|X_{\tilde{0}\tilde{1}}|$ increases [Figs. 3(o)].

## VI. Analytical calculation of transition frequencies

The transition frequencies described thus far have been calculated by diagonalizing the Hamiltonian Eq. (2), and are in good agreement with the experimental data. As shown in Appendix B, however, some of the transition-frequency shifts for small $\beta$ can be calculated analytically when $\Delta \sim \chi/2$, in which the frequency of the two-photon drive is in resonance with the energy difference between the states $|0\rangle$ and $|2\rangle$. Namely, the frequency shifts of $\omega_{\tilde{1}\tilde{0}}, \omega_{\tilde{0}\tilde{1}}, \omega_{\tilde{1}\tilde{2}},$ and $\omega_{\tilde{2}\tilde{1}}$ when $\Delta/2\pi = 8.20$ MHz are given by $+\left(\Delta + \frac{3\beta^2}{\chi}\right) - \sqrt{2}\beta, -\left(\Delta + \frac{3\beta^2}{\chi}\right) + \sqrt{2}\beta, +\left(\Delta + \frac{3\beta^2}{\chi}\right) + \sqrt{2}\beta$ and $-\left(\Delta + \frac{3\beta^2}{\chi}\right) - \sqrt{2}\beta$, respectively, where the term $\frac{3\beta^2}{\chi}$ is caused by the Stark shift of the state $|1\rangle$ and the term $\sqrt{2}\beta$ results from the Rabi splitting originating from the states $|\tilde{0}\rangle$ and $|\tilde{2}\rangle$. Figure 5 shows the comparison of transition frequencies obtained from diagonalizing the Hamiltonian (gray lines) and analytical formula (dashed lines). While they are in good agreement for small $\beta$, a discrepancy can be seen in the range of $\beta > 2\pi \times 2$ MHz especially for $\omega_{\tilde{1}\tilde{0}}$ and $\omega_{\tilde{0}\tilde{1}}$. This is because the contribution of the states with large photon numbers cannot be ignored when $\beta$ is large.

## VII. Comparison of nominal relaxation rates

Thus far, we have investigated the $\beta$ dependence of the frequency and the amplitude of peaks and dips in $|\Gamma|$ and found good agreement between the experiment and the theory as shown in Figs. 3(a)–3(f). However, the overall linewidth of the peaks and dips in the experiments is smaller than the theoretical as seen in the figures, where it should be noted that the color scales are all the same. To examine this difference, we investigate the nominal loss rates $\kappa_e^{(\tilde{m}\tilde{n})}$ and $\kappa_i^{(\tilde{m}\tilde{n})}$ obtained by fitting the measured $|\Gamma|$ to $\Gamma_{mn}$ in Eq. (6).

By comparing Eqs. (3) and (6), the nominal external loss rate $\tilde{\kappa}_e^{(\tilde{m}\tilde{n})}$ and internal loss rate $\tilde{\kappa}_i^{(\tilde{m}\tilde{n})}$ are given by

$$\tilde{\kappa}_e^{(\tilde{m}\tilde{n})} = -\kappa_e |X_{\tilde{m}\tilde{n}}|^2 (\rho_{\tilde{m}\tilde{m}} - \rho_{\tilde{n}\tilde{n}}), \tag{7}$$

$$\tilde{\kappa}_i^{(\tilde{m}\tilde{n})} = (\kappa_e + \kappa_i)(Y_{\tilde{m}\tilde{m}} + Y_{\tilde{n}\tilde{n}}) + \kappa_e |X_{\tilde{m}\tilde{n}}|^2 (\rho_{\tilde{m}\tilde{m}} - \rho_{\tilde{n}\tilde{n}}). \tag{8}$$

In Fig. 6, we plot $\kappa_e^{(\tilde{m}\tilde{n})}$ and $\kappa_i^{(\tilde{m}\tilde{n})}$ for $\Delta/2\pi = +8.20$ MHz as a function of $\beta$. For comparison, we also plot those predicted from the model, $\tilde{\kappa}_e^{(\tilde{m}\tilde{n})}$ and $\tilde{\kappa}_i^{(\tilde{m}\tilde{n})}$ [Eqs. (7) and (8)]. As shown in Fig. 6(a), $\kappa_e^{(\tilde{m}\tilde{n})}$ agrees well with $\tilde{\kappa}_e^{(\tilde{m}\tilde{n})}$ for all $(\tilde{m}, \tilde{n})$, except that $\kappa_e^{(\tilde{3}\tilde{0})}$ differs significantly from $\tilde{\kappa}_e^{(\tilde{3}\tilde{0})}$ in the range of $\beta/2\pi$ from 0.4 MHz to 4 MHz. This deviation is due to the fact that the dips in the transitions $|\tilde{0}\rangle \to |\tilde{3}\rangle$ at $\omega_{\tilde{3}\tilde{0}}$ and $|\tilde{2}\rangle \to |\tilde{3}\rangle$ at $\omega_{\tilde{3}\tilde{2}}$ overlap with each other in this pump power range [Fig. 3(d)]. Although the frequency difference between $\omega_{\tilde{3}\tilde{0}}$ and $\omega_{\tilde{3}\tilde{2}}$ is also small in the region of $\beta/2\pi < 0.4$ MHz, the deviation is not seen because the population difference between $\rho_{\widetilde{2}\widetilde{2}}$ and $\rho_{\widetilde{0}\widetilde{0}}$ makes the transition $|\widetilde{0}\rangle \to |\widetilde{3}\rangle$ dominant [Fig. 3(j)].

In contrast, the agreement between $\kappa_i^{(\widetilde{m}\widetilde{n})}$ and $\widetilde{\kappa}_i^{(\widetilde{m}\widetilde{n})}$ is poor as shown in Fig. 6(b). Only $\kappa_i^{(\widetilde{1}\widetilde{0})}$ and $\kappa_i^{(\widetilde{3}\widetilde{0})}$ are consistent with the theory in the limited range of $\beta/2\pi < 0.1$ MHz. Overall, $\widetilde{\kappa}_i^{(\widetilde{m}\widetilde{n})}$ is smaller than the theoretical estimation $\kappa_i^{(\widetilde{m}\widetilde{n})}$, except for $\kappa_i^{(\widetilde{3}\widetilde{0})}$ in the range of $\beta/2\pi$ from 0.4 MHz to 4 MHz, where we observed deviation in $\kappa_e^{(\widetilde{3}\widetilde{0})}$ as discussed above.

The discrepancy between $\widetilde{\kappa}_i^{(\widetilde{m}\widetilde{n})}$ and $\kappa_i^{(\widetilde{m}\widetilde{n})}$ is not well understood at this time. Because the nominal external photon loss rate, which is related to the single-photon decay to the signal line, is consistent with the theory, we believe that dephasing or losses other than the single-photon loss can cause the discrepancy between $\widetilde{\kappa}_i^{(\widetilde{m}\widetilde{n})}$ and $\kappa_i^{(\widetilde{m}\widetilde{n})}$. We do not include the phase relaxation effect in Eqs. (4) and (6), but we found that simply including phase relaxation in the master equation by $\mathcal{L} = a^\dagger a$ as the Lindblad operator does not reproduce $\widetilde{\kappa}_i^{(\widetilde{m}\widetilde{n})}$. One possibility is the $1/f$ spectrum of flux noise [31,32], which is not taken into account in the simulation. The phase relaxation and noise spectra will need to be futher investigated in future studies.

## VIII. Conclusion

We performed spectroscopic measurements of a KPO under the two-photon drive induced by the magnetic flux modulation and compared the results with theoretical calculations. The transition frequencies are in good agreement with the calculations after adjusting the attenuation values for the microwave transmission line, which turned out to be consistent with the results of independent measurements. We also demonstrated that some transition frequencies can be interpreted as Rabi splitting and the Stark shift and provided a simple analytical formula for them. The magnitude of the reflection coefficients is also in good agreement with the calculation, and we showed that its behavior under the two-photon drive, such as the change from the absorption dip to amplification peak, can be explained by the difference in the population between the initial and final states of the transition. We also investigated the nominal loss rates and found that the external loss is in good agreement with the theoretical calculation, while the internal loss rate shows significant deviation from the theory, which still needs clarification. By using these spectroscopic results, it is possible to precisely determine $\beta$, the magnitude of the two-photon drive, which is important for the precise control of the KPOs used as qubits in quantum annealing machines and quantum computers.

ACKNOWLEDGMENTS

We thank Y. Kitagawa for his assistance in the device fabrication. We also thank T. Aoki, K. Koshino, and Y. Kano for their fruitful discussions. A part of this work was conducted at the AIST Nano-Processing Facility supported by Nanotechnology Platform Program of the Ministry of Education, Culture, Sports, Science and Technology (MEXT), Japan. The devices were fabricated in the Superconducting Quantum Circuit Fabrication Facility (Qufab) in the National Institute of Advanced Industrial Science and Technology (AIST). This paper is based on the results obtained from a project, JPNP16007, commissioned by the New Energy and Industrial Technology Development Organization (NEDO). This work was partially supported by JST Moonshot R&D (Grant Number JPMJMS226C).

Appendix A: Two-photon drive amplitude

In this Appendix, we derive the formula for the amplitude of the two-photon drive applied to the KPO [Eq. (5)] based on its relation to the modulation amplitude of the resonance frequency.

First, we derive the relation between the amplitude of the frequency modulation $\delta\omega$ and the power of the two-photon drive. We assume the time dependence of the resonance frequency as $\omega_r(t) = \omega_r + \delta\omega \cos \omega_p t$, where $\omega_r$ is the static resonance frequency and $\omega_p$ is the frequency of the two-photon drive. Then, $\delta\omega$ is related to the derivative of the resonance frequency with respect to the bias current $\frac{d\omega_r}{di}$ as $\delta\omega = \frac{d\omega_r}{di} i_{amp}$, where $i_{amp}$ is the amplitude of the ac current induced by the two-photon drive. Here, we assume that the mutual inductance between the pump line and the SQUID is frequency independent. Because the RMS value of the current $|i|$ is $|i| = 1/\sqrt{2} i_{amp}$, the average power $\bar{p}$ is given by $\bar{p} = Z_0 |i|^2 = \frac{Z_0 i_{amp}^2}{2}$, where $Z_0$ is a characteristic impedance of the pump line. Therefore, $\bar{p}$ (in dBm) can be written as $\bar{p} = 10 \log_{10} \left[ \frac{Z_0 i_{amp}^2 \times 1000}{2} \right]$, i.e., $i_{amp} = \sqrt{\frac{2}{Z_0 \times 1000}} 10^{\frac{\bar{p}}{20}}$, and $\delta\omega$ is written as

$$\delta\omega = \left| \frac{d\omega}{di} \right| \sqrt{\frac{2}{Z_0 \times 1000}} 10^{\frac{\bar{p}}{20}}. \tag{A1}$$

The relation between the two-photon drive $\beta$ and frequency-modulation amplitude can be determined from the Hamiltonian Eq. (2) in a laboratory frame,

$$\frac{\mathcal{H}}{\hbar} = \omega_0 a^\dagger a - \frac{\chi}{12}(a^\dagger + a)^4 + 2\beta \cos(\omega_p t)(a^\dagger + a)^2. \tag{A2}$$

By expanding and rearranging the right-hand side of this equation into normally ordered products, the coefficient of $a^\dagger a$ becomes $\omega_0 - \chi + 4\beta \cos(\omega_p t)$. The constant terms are the resonance frequency $\omega_r = \omega_0 - \chi$ of the KPO without the two-photon drive, and the oscillating term represents the frequency modulation caused by the two-photon drive. Therefore, the frequency-modulation amplitude $\delta\omega$ can be written as $\delta\omega = 4\beta$, which together with Eq. (A1) leads to Eq. (5).

Appendix B: Analytical calculation of frequency shifts

In this Appendix, we derive formulae for four of the transition frequencies as a function of the two-photon drive amplitude $\beta$, which are discussed in Section VI and shown in Fig. 5. We investigate the physical meaning of the frequency shift by deriving its formula analytically.

We consider the situation where $\omega_p$ matches the energy difference between $|0\rangle$ and $|2\rangle$. In this case, the two-photon drive acts as a resonant drive for $|0\rangle$ and $|2\rangle$ and as a detuned drive for the states with two different photon numbers, e.g., for $|1\rangle$ and $|3\rangle$ [Fig. 7(a)]. It is well known that a resonant drive applied to a two-level system causes Rabi splitting and a non-resonant drive causes the Stark shift [2]; here we extend the same idea to the two-photon drive applied to the KPO.

If the states with a photon number of less than 3 are considered, the Hamiltonian Eq. (2) is rewritten as

$$\frac{\mathcal{H}_{(0-3)}}{\hbar} = \sqrt{2}\beta(|0\rangle\langle 2| + |2\rangle\langle 0|)$$
$$+\sqrt{6}\beta(|1\rangle\langle 3| + |3\rangle\langle 1|) + \frac{1}{2}\chi|1\rangle\langle 1| - \frac{3}{2}\chi|3\rangle\langle 3|. \tag{A3}$$

We decompose $\mathcal{H}_{\text{RWA}}$ as

$$\mathcal{H}_{(0-3)} = \mathcal{H}_{\text{Rabi}} + \mathcal{H}_{\text{Stark}}, \tag{A4}$$

where

$$\frac{\mathcal{H}_{\text{Rabi}}}{\hbar} = \sqrt{2}\beta(|0\rangle\langle 2| + |2\rangle\langle 0|), \tag{A5}$$

$$\frac{\mathcal{H}_{\text{Stark}}}{\hbar} = \sqrt{6}\beta(|1\rangle\langle 3| + |3\rangle\langle 1|) + \frac{1}{2}\chi|1\rangle\langle 1| - \frac{3}{2}\chi|3\rangle\langle 3|. \tag{A6}$$

$\mathcal{H}_{\text{Rabi}}$ represents a resonant drive for $|0\rangle$ and $|2\rangle$ states, and $\mathcal{H}_{\text{Stark}}$ represents a detuned drive for $|1\rangle$ and $|3\rangle$ states (Fig. 7a).

First, we focus on $\mathcal{H}_{\text{Rabi}}$. This term is described as a two-level system of $|0\rangle$ and $|2\rangle$. The diagonalization of $\mathcal{H}_{\text{Rabi}}$ yields

$$\frac{\mathcal{H}_{\text{Rabi}}}{\hbar} = \sqrt{2}\beta(|+_2\rangle\langle +_2| - |-_2\rangle\langle -_2|). \tag{A7}$$

Here, we define $|\pm_2\rangle = \frac{1}{\sqrt{2}}(|2\rangle \pm |0\rangle)$. This is analogous to the derivation of the Rabi splitting for two-level systems, which can be observed as a Mollow triplet.

Next, we focus on $\mathcal{H}_{\text{Stark}}$. This term is described as a two-level system of $|1\rangle$ and $|3\rangle$ with a detuned drive. The diagonalization of $\mathcal{H}_{\text{Stark}}$ yields

$$\frac{\mathcal{H}_{\text{Stark}}}{\hbar} = \left(-\frac{1}{2}\chi + \chi\sqrt{1+\frac{6\beta^2}{\chi^2}}\right)|1'\rangle\langle 1'| + \left(-\frac{1}{2}\chi - \chi\sqrt{1+\frac{6\beta^2}{\chi^2}}\right)|3'\rangle\langle 3'|, \quad (A8)$$

where

$$|1'\rangle = \frac{1}{\sqrt{1+\delta^2}}(|1\rangle - \delta|3\rangle), \quad (A9)$$

$$|3'\rangle = \frac{1}{\sqrt{1+\delta^2}}(\delta|1\rangle + |3\rangle). \quad (A10)$$

Here, we define $\delta = \frac{\chi-\sqrt{\chi^2+6\beta^2}}{\sqrt{6}\beta}$. If the amplitude of the two-photon drive is much smaller than the Kerr nonlinearity, i.e., $\left(\frac{\beta}{\chi}\right)^2 \ll 1$, Eq. (A8) is written as

$$\frac{\mathcal{H}_{\text{Stark}}}{\hbar} \simeq \left(\frac{1}{2}\chi + \frac{3\beta^2}{\chi}\right)|1'\rangle\langle 1'| + \left(-\frac{3}{2}\chi - \frac{3\beta^2}{\chi}\right)|3'\rangle\langle 3'|. \quad (A11)$$

Comparing Eqs. (A6) and (A11), it can be seen that the ac Stark shift is caused by the two-photon drive. Namely, the eigenenergies of $|1\rangle$ and $|3\rangle$ are shifted $+\frac{3\beta^2}{\chi}$ and $-\frac{3\beta^2}{\chi}$, respectively. From Eqs. (A7) and (A11), $\mathcal{H}_{\text{RWA}}$ can be written as

$$\begin{aligned}\frac{\mathcal{H}_{(0-3)}}{\hbar} &\simeq \sqrt{2}\beta(|+_2\rangle\langle +_2| - |-_2\rangle\langle -_2|) \\ &+ \left(\frac{1}{2}\chi + \frac{3\beta^2}{\chi}\right)|1'\rangle\langle 1'| + \left(-\frac{3}{2}\chi - \frac{3\beta^2}{\chi}\right)|3'\rangle\langle 3'|.\end{aligned} \quad (A12)$$

This rewriting from Eq. (A3) to Eq. (A12) corresponds to the rewriting of the Fock states to the dressed states shown as the orange and pink levels in Fig. 7(b), which are induced by the two-photon drive shown by the vertical arrows in their corresponding colors in Fig. 7(a).

Transition frequencies between the dressed states can be calculated from Fig. 7(b). Namely, frequencies for the transitions $|\pm, N\rangle \leftrightarrow |1, N\rangle$ (dark blue dashed arrows) are $\frac{\omega_p}{2} + \left(\frac{\chi}{2} + \frac{3\beta^2}{\chi}\right) \pm \sqrt{2}\beta$ and those for the transitions $|1, N\rangle \leftrightarrow |\pm, N\rangle$ (dark red dashed arrows) are $\frac{\omega_p}{2} - \left(\frac{\chi}{2} + \frac{3\beta^2}{\chi}\right) \pm \sqrt{2}\beta$. They are plotted in Fig. 5 in their corresponding colors.

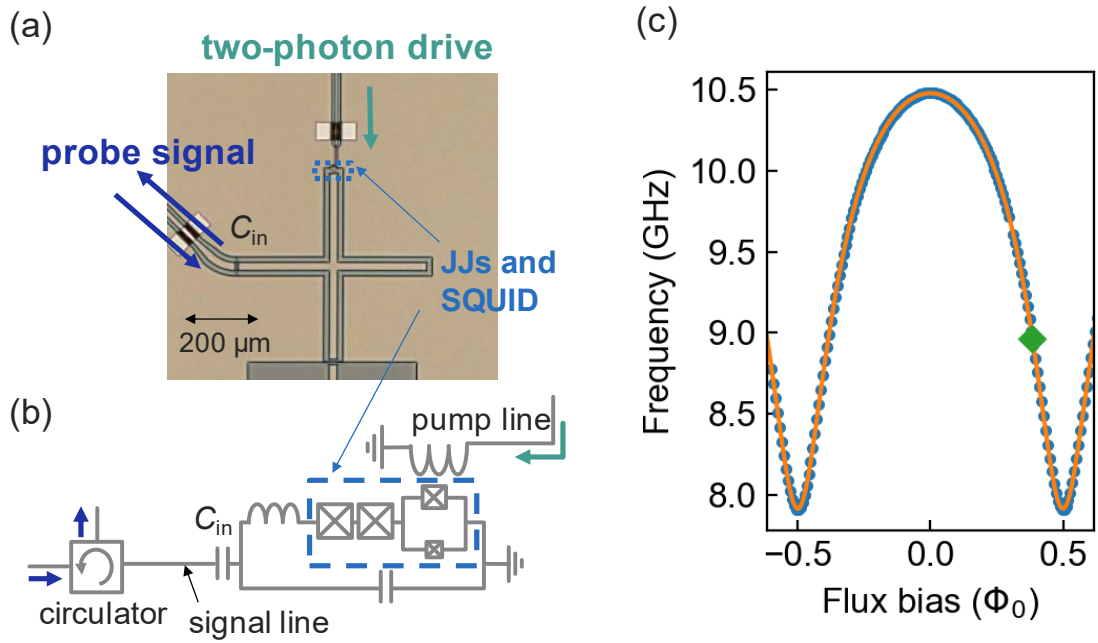

FIG. 1. (a) Optical micrograph of the KPO and (b) its schematic circuit diagram. The probe signal is injected from the signal line, which is capacitively coupled with the KPO. The reflected signal is separated from the injected signal by a circulator. The two-photon drive is applied from the pump line, which is inductively coupled to the SQUID in the KPO. The DC current is also applied to the pump line to induce a static magnetic field in the SQUID loop. (c) Flux-bias dependence of the resonance frequency. The blue dots are experimental data, and the orange line shows the theoretical calculation of the resonance frequency based on the circuit model shown in (b) and fitted to the data. The green diamond represents the operation point used for the spectroscopy measurements with the two-photon drive.

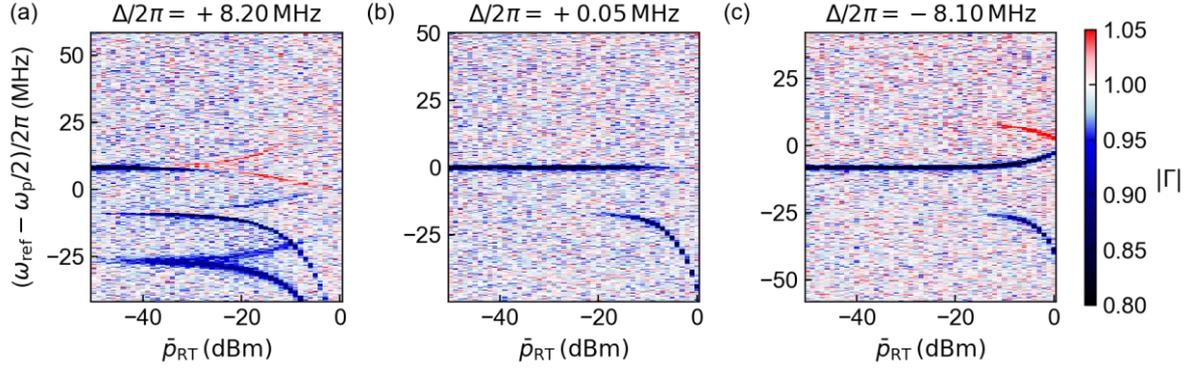

**FIG. 2. Amplitude of reflection coefficient as a function of the probe frequency $\omega_{\mathrm{ref}}$ and the two-photon drive power $\bar{p}_{RT}$ with different detunings of (a) $\Delta/2\pi = +8.20\,\mathrm{MHz}$, (b) $\Delta/2\pi = +0.05\,\mathrm{MHz}$, and (c) $\Delta/2\pi = -8.10\,\mathrm{MHz}$. The power of the two-photon drive $\bar{p}_{RT}$ is defined as that at the output of the signal generator.**

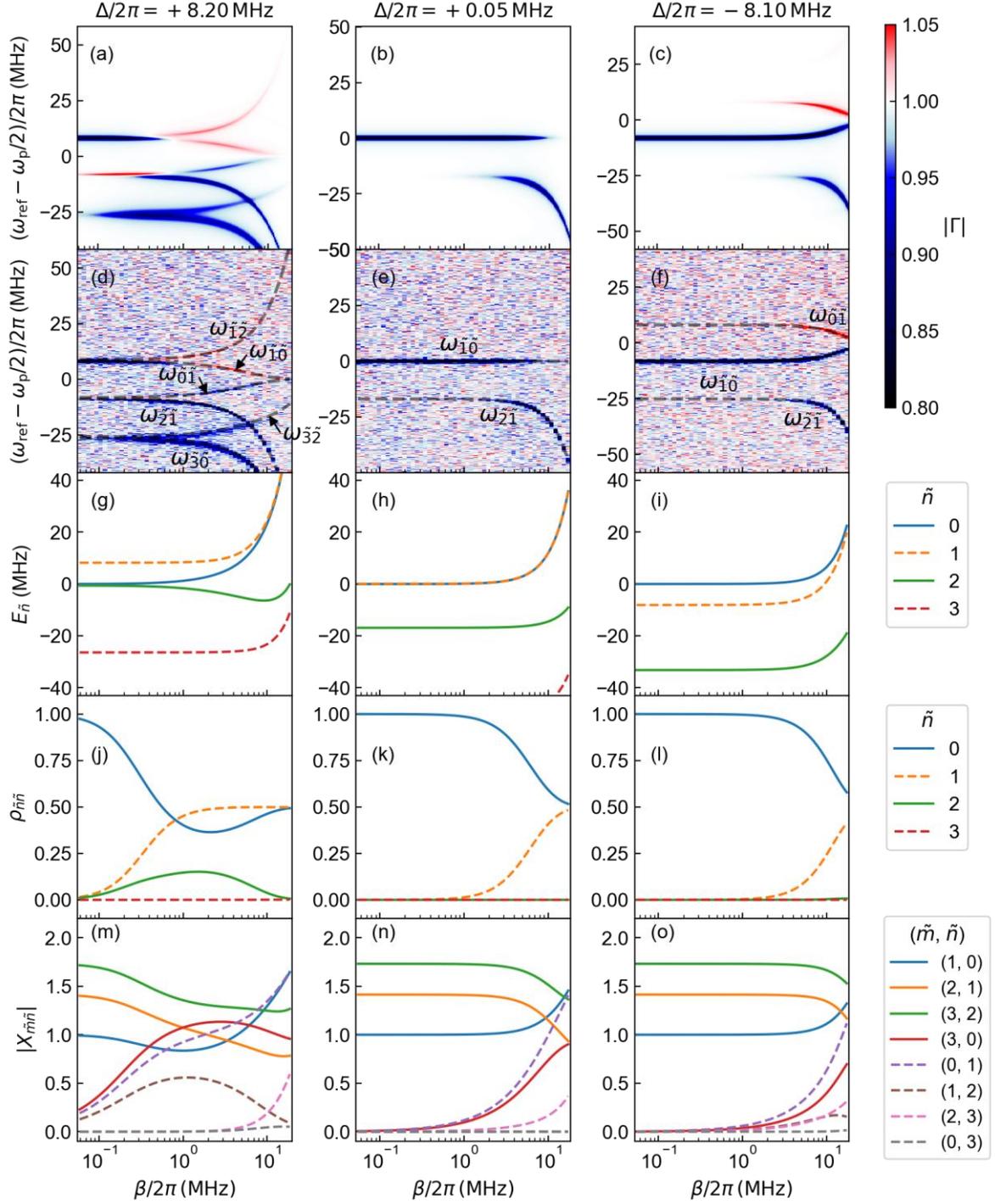

FIG. 3. Simulated amplitude of reflection coefficient as a function of the probe frequency $\omega_{\mathrm{ref}}$ and the amplitude of the two-photon drive $\beta$ with a detuning of (a) $\Delta/2\pi = +8.20$ MHz, (b) $\Delta/2\pi = +0.05$ MHz, and (c) $\Delta/2\pi = -8.10$ MHz. [(d)–(f)] Spectra in Fig. 2 replotted as a function of $\beta$ in the horizontal axis. [(g)–(i)] Energy of the eigenstates, [(j)–(l)] population of the energy eigenstates for the steady state calculated by a GKSL master equation, and [(m)–(o)] matrix elements of the one-photon transition as a function of two-photon drive amplitude $\beta$ with a different detuning of

$\Delta/2\pi = +8.20$ MHz [(g), (j), and (m)], $0.05$ MHz [(h), (k), and (n)], and $-8.10$ MHz [(i), (l), and (o)]. Gray dashed lines in (d)–(f) represent the transition frequencies calculated from eigenenergies shown in (g)–(i), respectively. Attenuations in the pump line $R_{\text{RT-JPO}}$ were determined to be (d) -57.5 dB, (e) -57.4 dB, and (f) -57.0 dB to minimize the square of the difference between the measured and calculated transition frequencies.

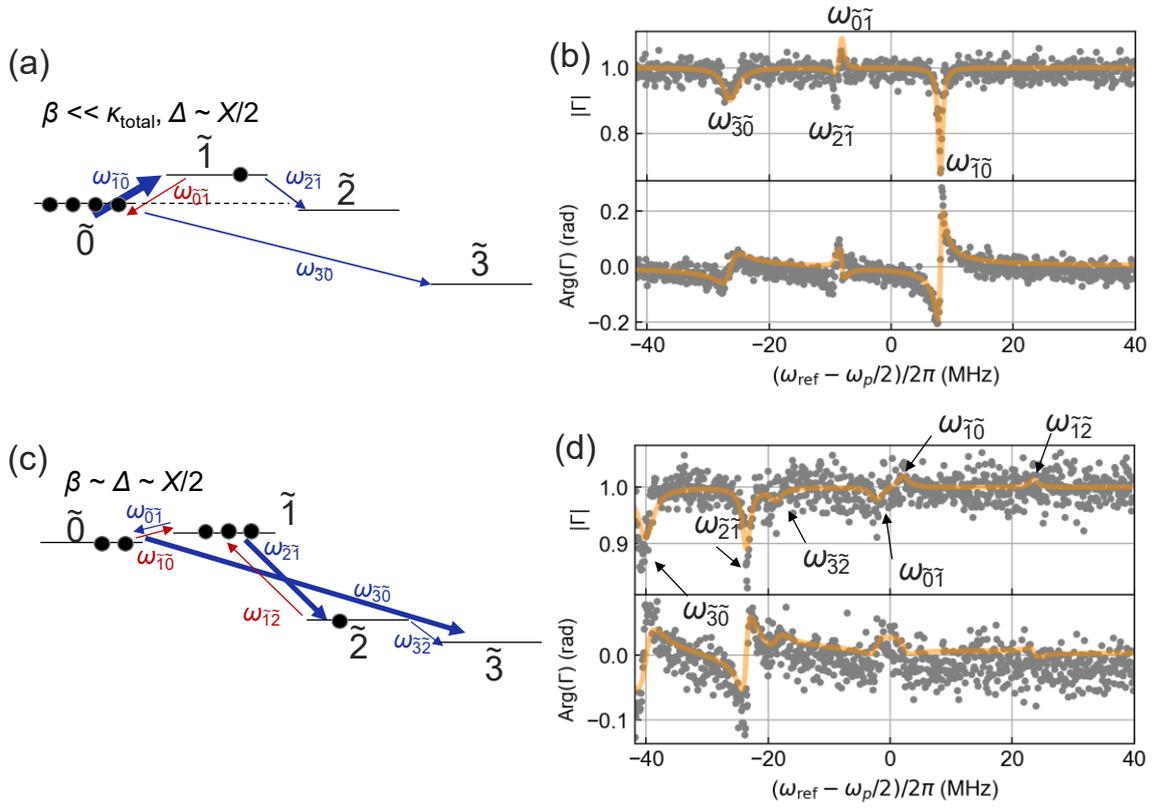

FIG. 4. Schematic energy diagram and the reflection coefficient as a function of the probe frequency for $\beta/2\pi = 0.21$ MHz (a, b) and $\beta/2\pi = 7.40$ MHz (c, d). In (a) and (c), the black dots in each energy levels represent the population of the corresponding state. Blue (red) arrows in the energy diagrams represent the transitions from the state with larger (smaller) population to the state with smaller (larger) population, which results in the absorption dips (amplification peaks) in $|\Gamma|$. Thickness of the arrows qualitatively represents the strength of the transition. In (b) and (d), gray dots represent the experimental data, while the solid orange curves represent the numerical simulations.

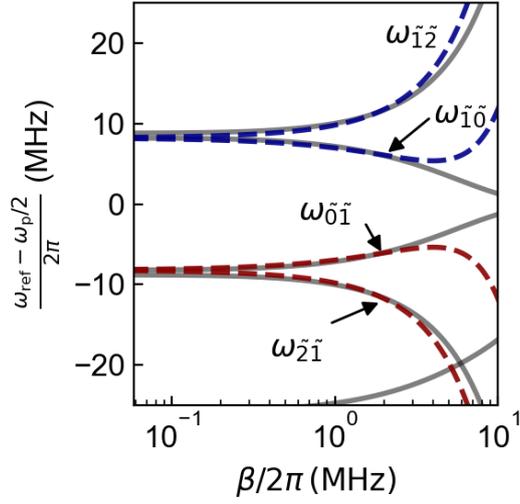

**FIG. 5.** Comparison of transition frequencies obtained from diagonalization of the Hamiltonian (gray solid lines) and analytical calculation for the detuning $\Delta \sim \chi/2$. Analytical calculation is shown by the dark blue and dark red dashed lines, which are given by $+\left(\Delta + \frac{3\beta^2}{\chi}\right) \pm \sqrt{2}\beta$ and $-\left(\Delta + \frac{3\beta^2}{\chi}\right) \pm \sqrt{2}\beta$, respectively.

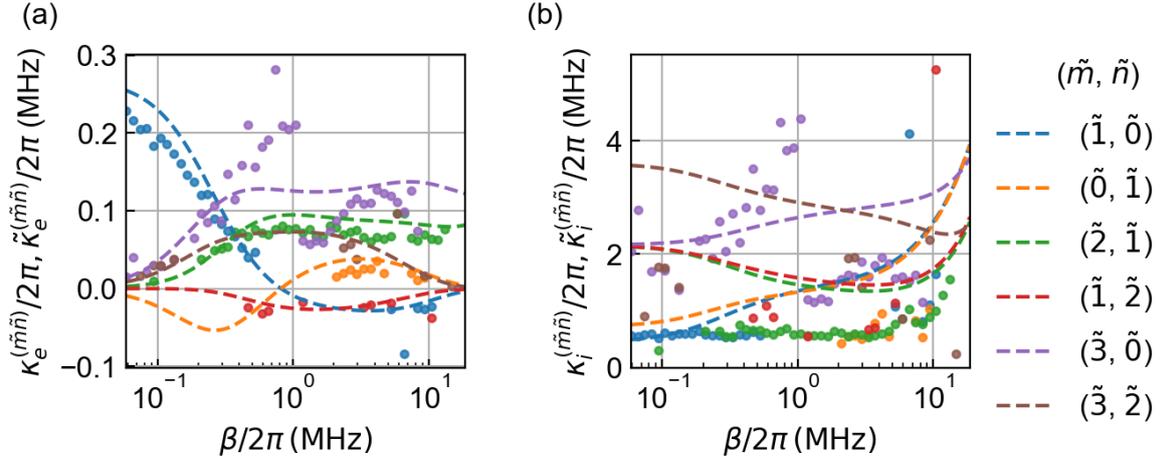

FIG. 6. Nominal (a) external and (b) internal photon loss rates as a function of two-photon drive amplitude at $\Delta/2\pi = +8.20$ MHz. The dots represent the experimental data [$\kappa_e^{(\tilde{m}\tilde{n})}$ and $\kappa_i^{(\tilde{m}\tilde{n})}$], and the dashed lines represent the calculation [$\tilde{\kappa}_e^{(\tilde{m}\tilde{n})}$ and $\tilde{\kappa}_i^{(\tilde{m}\tilde{n})}$].

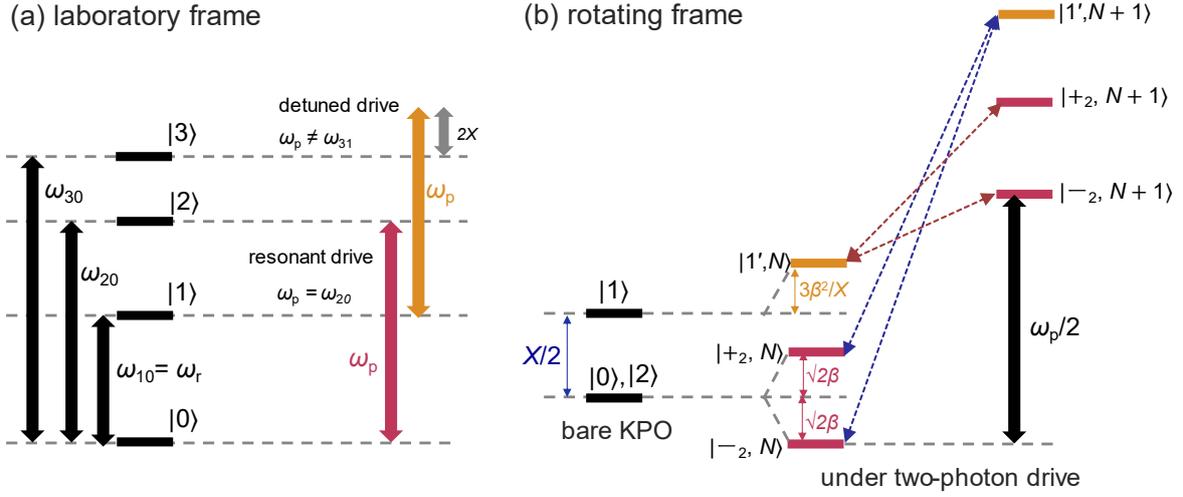

FIG. 7. Schematic energy diagram under the condition that the parametric drive frequency $\omega_p$ is resonant to the energy difference between the states $|0\rangle$ and $|2\rangle$. (a) Energy-level diagram in a laboratory frame. The energy level is represented by Fock states and $\omega_{nm}$ is the energy difference between the $n$-th and $m$-th Fock states. The parametric drive works in two different ways; the resonant drive for the states $|0\rangle$ and $|2\rangle$ (pink arrow) and the detuned drive for the states $|1\rangle$ and $|3\rangle$ (orange arrow). (b) Energy-level diagram in a rotating frame and possible state transitions. Left: the bare KPO energy levels. Center and right: the energy levels with a photon number of $N$ and $(N+1)$ under the two-photon drive, respectively. The resonant drive between the states $|0\rangle$ and $|2\rangle$ causes the Rabi splitting (energy levels in pink). The detuned drive between the states $|1\rangle$ and $|3\rangle$ changes the energy of the state $|1\rangle$ (energy level in orange) by ac Stark effect. The dark blue and dark red dashed arrows represent the transitions induced by the probe signals.